\documentclass[bibyear,usenatbib]{aa}  

\usepackage{graphicx}
\usepackage{txfonts}
\usepackage {natbib}

\begin{document}

   \title{CO luminosity - line width correlation of sub-millimeter galaxies}
   \titlerunning{CO luminosity - line width correlation of sub-millimeter galaxies}
   \authorrunning{GOTO \& TOFT}

   \subtitle{and a possible cosmological application}

   \author{Tomotsugu Goto
          \inst{1,2}
          \and
          Sune Toft\inst{2}\fnmsep
%\thanks{Just to show the usage          of the elements in the author field}
          }

   \institute{National Tsing Hua University, No. 101, Section 2, Kuang-Fu Road, Hsinchu, Taiwan 30013\\
              \email{tomo@phys.nthu.edu.tw}
         \and
	 Dark Cosmology Centre, Niels Bohr Institute, University of Copenhagen, Denmark\\
             \email{sune@dark-cosmology.dk}
 %            \thanks{The university of heaven temporarily does not                   accept e-mails}
             }

   \date{Received March 11, 2015; accepted March 11, 2015}

% \abstract{}{}{}{}{} 
% 5 {} token are mandatory
 
  \abstract
   {
A possible correlation between CO luminosity ($L'_{CO}$) and its line width (FWHM) has been suggested and denied in the literature. Such claims were often based on a small, or heterogeneous sample of galaxies, and thus inconclusive.
}
   {
We aim to prove or dis-prove the $L'_{CO}$-FWHM correlation.
}
   {
We compile a large sample of submm galaxies at $z>2$ from the literature, and investigate the $L'CO$-FWHM relation.
   }
   {
 After carefully evaluating the selection effects and uncertainties such as inclination and magnification via gravitational lensing, we show that there exist a weak but significant correlation between $L'_{CO}$ and FWHM. We also discuss a feasibility to measure the cosmological distance using the correlation. 
}
   {}


   \maketitle
%
%________________________________________________________________

\section{Introduction}

A possible correlation between CO luminosity ($L'_{CO}$) and its line width at high-z has been suggested in the literature.
\citet{2013MNRAS.429.3047B} observed 32 submm galaxies (SMG) at $1.2<z<4.1$ in $^{12}$CO and suggested a correlation between CO luminosity and line width (FWHM) in high-redshift starbursts, likely relating to baryon-dominated gas dynamics within the regions probed (See their Fig. 5). 
Using 15 SMG at z=2-4 with known lens magnifications, \citet{2012ApJ...752..152H} showed that the SMGs fall close to a power-law with small scatter in CO luminosity and line width (FWHM) plane (See their Fig.7).  

 Bright SMGs are often gravitationally-lensed. Such relation between $L'_{CO}$ and FWHM can be used to estimate magnification factor ($\mu$). Indeed, \citet{2012ApJ...752..152H} used the relation to estimate luminosity of SMGs with unknown $\mu$. 
 \citet{2013Natur.498..338F} also used this relation to estimate a small $\mu$ ($\sim$1.8) of merging galaxies at z=2.3 to conclude their star formation rate were 2000 $M_{\odot} yr^{-1}$ (See their Fig. S7).
Similarly, \citet{2013Natur.496..329R}  used the relation to estimate $\mu$ (1.5$\pm$0.7) of a starburst galaxy at z=6.34, finding its SFR of 2900 $M_{\odot} yr^{-1}$.

 Conversely, \citet{2013ARA&A..51..105C} examined the relation using a much larger sample, and found no significant correlation between  $L'_{CO}$ and FWHM (See their Fig.5). However, their sample was heterogeneous including various types of galaxies such as SMM, QSO, radio galaxies, 24$\mu$m sources and so on. 
 Therefore, the existence of $L'_{CO}$-FWHM correlation is still controversial.

In this paper, we compile a larger sample of $z>2$ galaxies with CO detection from the literature, to re-examine the  $L'_{CO}$-FWHM correlation at high-z. We carefully examine various sources of possible uncertainties such as different types of galaxies, luminosity conversion from different CO transitions, and errors in estimating $\mu$.
  Unless otherwise stated, we adopt a cosmology with $(h,\Omega_m,\Omega_\Lambda) = (0.7,0.3,0.7)$.% \citep{2008arXiv0803.0547K}. 

\begin{figure}
\begin{center}
\includegraphics[scale=0.65]{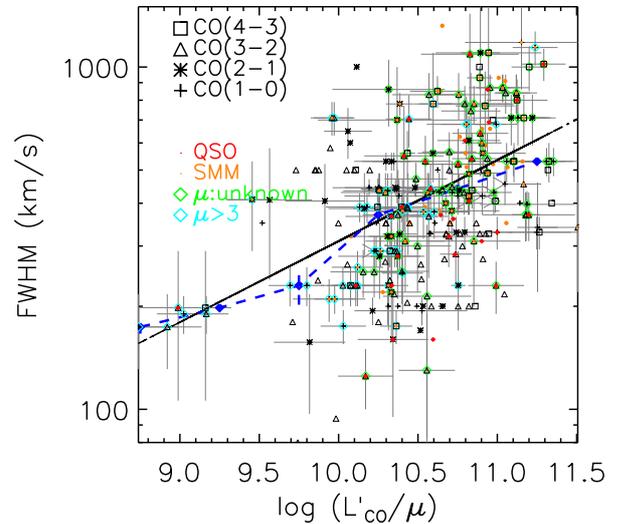}
\end{center}
\caption{
 Correlation between $L'_{CO}/\mu$ and FWHM for various samples. The blue dotted line connects the median in equally spaced logarithmic bins. The black solid line is the best-fit power-law.
}\label{fig:correlation}
\end{figure}

\begin{table*}
 \centering
 \begin{minipage}{180mm}
  \caption{Correlation between $L'_{CO}/\mu$ and FWHM for various samples. Dispersion is around the best-fit power-law. $p$-value of $<$0.05 can be considered significant.}\label{tab:corr_coeff}
  \begin{tabular}{@{}ccclllcccc@{}}
  \hline
Sample &  N &Spearman's coeff. &p-value&Kendall's coeff.&p-value & Dispersion (km/s) \\ 
 \hline
   \hline
%All &          169 & 0.54 & 2.4e-14 & 0.39 & 0.0e+00 & 194.\\
All  &          195 & 0.55 & 1.7$\times10^{-16}$ & 0.39 & $<1.0\times10^{-16}$ & 187.\\
   \hline
%Submm glaxies &           46 & 0.57 & 3.9e-05 & 0.40 & 8.9e-05 & 244.\\
Submm galaxies  &           72 & 0.53 & 2.1$\times10^{-6}$ & 0.37 & 4.2$\times10^{-6}$ & 272.\\
%QSOs &           31 & 0.53 & 2.0e-03 & 0.39 & 1.9e-03 & 245.\\
QSO hosts &           31 & 0.53 & 2.0$\times10^{-3}$ & 0.39 & 1.9$\times10^{-3}$ & 245.\\
%Submm glaxies +QSO &           77 & 0.54 & 3.6e-07 & 0.39 & 6.0e-07 & 236.\\
Submm galaxies +QSO hosts &          103 & 0.52 & 1.4$\times10^{-8}$ & 0.37 & $<1.0\times10^{-16}$ & 218.\\%Bothwell
   \hline
CO(4-3) only &           26 & -0.34 & 9.2$\times10^{-2}$ & -0.21 & 1.4$\times10^{-1}$ & 279.\\
CO(3-2) only &           62 & 0.12 & 3.5$\times10^{-1}$ & 0.09 & 2.8$\times10^{-1}$ & 217.\\
CO(2-1) only &           38 & 0.18 & 2.8$\times10^{-1}$ & 0.14 & 2.2$\times10^{-1}$ & 182.\\
CO(1-0) only &           43 & 0.22 & 1.5$\times10^{-1}$ & 0.15 & 1.6$\times10^{-1}$ & 121.\\
   \hline
All but CO(4-3) &          143 & 0.52 & 4.0$\times10^{-11}$ & 0.37 & $<1.0\times10^{-16}$ & 206.\\
All but CO(3-2) &          107 & 0.56 & 2.5$\times10^{-10}$ & 0.40 & $<1.0\times10^{-16}$ & 201.\\
All but CO(2-1) &          131 & 0.58 & 5.9$\times10^{-13}$ & 0.42 & $<1.0\times10^{-16}$ & 231.\\
All but CO(1-0) &          126 & 0.52 & 4.8$\times10^{-10}$ & 0.37 & $<1.0\times10^{-16}$ & 237.\\
  \hline
%$\mu$ known &          111 & 0.49 & 3.9e-08 & 0.35 & 6.0e-08 & 156.\\%Bothwell
$\mu$ known &           85 & 0.42 & 7.3$\times10^{-5}$ & 0.30 & 5.3$\times10^{-5}$ & 156.\\
%$0<\mu<2$ &           31 & -0.14 & 4.6e-01 & -0.07 & 5.6e-01 & 161.\\
%$\mu$ unknown &           84 & 0.46 & 1.2e-05 & 0.32 & 2.1e-05 & 243.\\
$\mu$ unknown &          110 & 0.47 & 2.4$\times10^{-7}$ & 0.33 & 3.6$\times10^{-7}$ & 205.\\ %(Bothwell)
   \hline
24$\mu$m sources &            8 & 0.45 & 2.6$\times10^{-1}$ & 0.36 & 2.2$\times10^{-1}$ & 204.\\
%Radio glaxies &            4 & 0.20 & 8.0e-01 & 0.00 & 1.0e+00 & 249.\\
BzK galaxies &            4 & 0.80 & 2.0$\times10^{-1}$ & 0.67 & 1.7$\times10^{-1}$ & 156.\\
% Compiled module: LEGEND.
 \hline
\end{tabular}
\end{minipage}
\end{table*}

\section{Sample}
\label{Data}

Our sample is based on updated version of the supplemental table of \citet{2013ARA&A..51..105C}, 
which contains all published CO line measurements at $z>2$ by December 31, 2012. 
The updated version also include some submm sources published in 2013.
We created a large sample of high-z submm galaxies by adding 32 sources observed by \citet{2013MNRAS.429.3047B}.
In total, our sample contains 195 galaxies, detected at least in one CO transition, with a measured FWHM.

\section{Analysis}\label{sec:ana}

 We compute $L'_{CO}$ using the standard definition  
\begin{equation} \label{eqn:lumfunc2p}
 L_{CO}'=3.25 \times 10^7 S_{CO} \Delta v \nu_{obs}^{-2} D_L^2 (1+z)^{-3},
\end{equation}
where  $S_{CO} \Delta v$ is the velocity integrated line flux, $L_{CO}'$ is the line luminosity in K Km s$^{-1}$ pc$^2$, $\nu_{obs}$ is the observed central frequency of the line and $D_L$ is the luminosity distance in Mpc.

If the molecular gas emission comes from thermalized, optically thick regions,  $L'_{CO}$ is constant for all $J$ levels. 
However, in reality, the ratio of $L_{CO}'(1-0)$ to higher order $L_{C0}'(J,J-1)$ depends on temperature and density.  
 We use average ratio of $L_{CO}'(J,J-1)$/$L_{CO}'(1-0)$ provided in Table 4 of \citet{2013MNRAS.429.3047B}, to calculate $L_{CO}'(1-0)$ from higher order CO observations. Higher redshift galaxies are often observed only at mid-to-high-J CO transitions.
 We discuss uncertainties associated with this conversion in Section \ref{sec:results} in detail.

It is always a concern with high-z objects, that they might be gravitationally magnified. Submm galaxies are not an exception.  $L_{CO}'$ thus also needs to be corrected for a magnification factor ($\mu$), when the galaxy is magnified by a gravitational lens. We collected $\mu$-values from the literature whenever available. However, as $\mu$ must be estimated from detailed modeling of the lens potential, a reliable measurement is not always available. We will examine how the correlation depends on the availability and amplitude of  $\mu$ in Section \ref{sec:results}, but in brief, they do not change our main results. Most likely, majority of galaxies with missing $\mu$ are not strongly magnified; otherwise a lens galaxy would have been detected and $\mu$ would have been estimated. 
 We present  $L'_{CO}/\mu$, whenever $\mu$ is available hereafter.

\section{Results}\label{sec:results}
\subsection{$L'_{CO}/\mu$-FWHM correlation}

In Fig.1, we plot CO line width FWHM against $L'_{CO}/\mu$ for all galaxies in our sample. 
The solid line is the best-fit power-law, which is 
%;FWHM (km/s) =(231$\pm$31) $\times$ log$_{10}$($L'_{CO}/\mu$) + 1970$\pm$330. 
%FWHM (km/s) $\propto$$L'_{CO}/\mu^{0.24\pm0.03}$.
%\begin{equation}\label{equation}
%FWHM (km/s) = (1.3\pm0.3)\times (L'_{CO}/\mu)^{0.24\pm0.03}
%\end{equation}
\begin{equation}\label{bestfit}
\log (FWHM) =   (0.24\pm0.03) \times \log(L'_{CO}/\mu) + (0.12\pm0.28)
\end{equation}
%y=(       1.3266203+-       1.9421681)Lco^(      0.23685954+-     0.027298839)
%Significance of non-zero slope is        8.6765426
The significance of the non-zero slope is 8.7~$\sigma$. Spearman’s correlation coefficient is 0.55 with p-value of 1.7$\times$10$^{-16}$. Kendall’s correlation coefficient is similarly significant as shown in Table 1. The blue dashed line connects the median points in each bin. Errors on each point are small due to the large number of galaxies in our sample. The median values increase with increasing $L'_{CO}/\mu$. These results show there exists a weak, but very significant correlation between $L'_{CO}/\mu$ and FWHM.

The presence of such a correlation is not surprising. $L'_{CO}/\mu$ is sensitive to the total mass of molecular gas, while the FWHM is sensitive to dynamical mass. The relation is similar to Tully-Fisher relation, which is also a power-law, in the local Universe \citep{1977A&A....54..661T}. A large scatter around the relation is also expected due to the inclination effect, and the unknown $\mu$ for a subset of galaxies. We measured a dispersion around the best-fit line of 43\%, which is consistent with what one expects from randomly oriented thin disks (45\%), suggesting $\mu$ is close to 1 for most of the galaxies with unknown $\mu$.

\begin{figure}
\begin{center}
\includegraphics[scale=0.5]{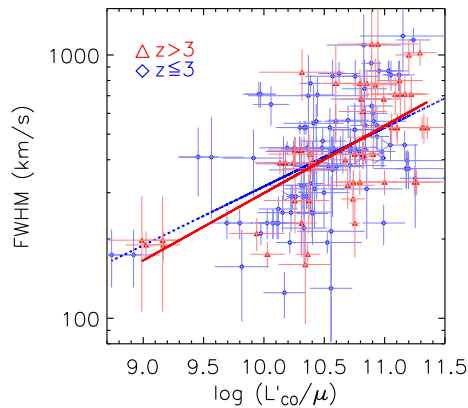}
\end{center}
\caption{
$L'_{CO}/\mu$ is plotted against FWHM. 
 We separated the sample into z$>$3 (red triangles) and z$\leq$3 (blue diamonds). The best-fit power laws for the two samples are;  
FWHM (km/s) $\propto$$L'_{CO}/\mu^{0.22\pm0.04}$ (z$\leq$3, blue)
, and 
FWHM (km/s) $\propto L'_{CO}/\mu^{0.25\pm0.04}$ (z$>$3, red). 
The slopes are consistent with each other, and with the slope derived from the full sample in Fig.1.
}\label{fig:corr_z}
\end{figure}

\begin{figure}
\begin{center}
\includegraphics[scale=0.5]{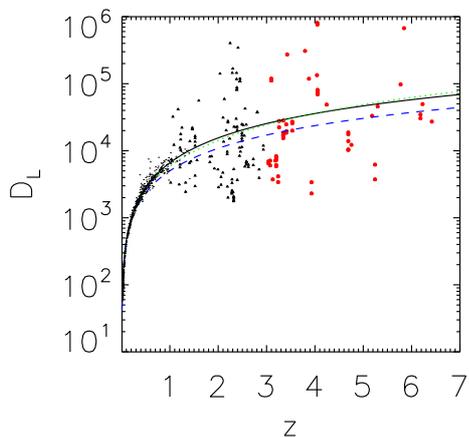}
\end{center}
\caption{
Luminosity distance in Mpc against z. 
The red circles and black triangles are our CO samples separated at z=3.
The black dots are 192 supernovae used to calibrate our low-z sample.
The black solid, green dotted and blue-dashed lines are $(\Omega_M,\Omega_\Lambda)$=(0.3,0.7), (0.3,0.0), and (1.0,0.0), respectively.
}\label{fig:co_hubble}
\end{figure}

\subsection{Uncertainty in converting different transitions}
We test if conversions from different CO transitions can mimic the correlation. In Fig.1,we used different symbols for different CO transitions (4-3,3-2,2-1, and 1-0). In Table 1, we show correlation coefficients for these. For CO(4-3), we do not see a positive correlation, while we see weak correlations for the other transitions. It is reassuring to see the correlations within single CO transitions, because they cannot be due to errors in the transition conversion. The reason why we do not see a positive correlation for the CO(4-3) transition is not clear, but we point out that this is the smallest subsample, and that since a single transition only spans a small redshift range (and thus, a small luminosity range), it is difficult to examine a correlation with a small sample in a single transition. It will be necessary to repeat this test using the wider frequency coverage of ALMA. On the other hand, we see significant correlations in all cases when we exclude a certain CO transition from the sample. These tests suggest that the $L'_{CO}$-FWHM correlation is not an artifact of CO order conversions.

\subsection{Dependence on magnification}
Among our sample, 85 galaxies have known $\mu$, while the other 110 galaxies do not. We should first note that many submm galaxies with unknown $\mu$ are probably not strongly magnified; otherwise we would have seen the lens galaxy/cluster and measured $\mu$. With this in mind, we still test if an unknown $\mu$ can mimic the correlation. 
In Fig.\ref{fig:correlation}, we mark two extreme samples of galaxies; galaxies with $\mu>3$ in cyan, and those with unknown $\mu$ in green. There is no apparent deviation from the best-fit relation, except that more $\mu$s are unknown at higher redshift. 
In Table 1, we separate the sample into those with known $\mu$, and those with unknown $\mu$, both of which show a significant correlation. Because brighter submm galaxies are more likely to be lensed (due to the steep luminosity function), the correction to unknown $\mu$ will steepen the correlation, i.e., the correlation will not disappear with the correction. It is re-assuring to see a positive correlation using many submm galaxies with unknown $\mu$.

\subsection{Dependence on galaxy types}
We also separate the sample according to galaxy types in  Table \ref{fig:correlation} and Fig.\ref{fig:correlation}.
Unfortunately, 24$\mu$m sources and BzK galaxy samples are too small to examine the correlation.
Both submm galaxies (SMG) and QSOs have a significant correlation, with submm galaxies having a greater significance.

\subsection{Redshift dependence}
Next we examine if the $L'_{CO}$-FWHM correlation changes with redshift. Unfortunately, due to the flux limits of the telescopes, the higher redshift sample includes more luminous sources, while lower redshift sources have smaller luminosity due to volume effects. With this limitation in mind, we separated the sample into high (z$>$3) and low (z$\leq$3) redshift samples in Fig.2. The best-fit lines to each sample agree with each other within the errors, consistent with no significant redshift-evolution. The slopes are also consistent with each other, and with that of the full sample in Fig.1.

A concern for low level CO transition is that CMB temperature becomes higher at the highest redshift end. For example, the CMB temperature at z=6 is $\sim$19K, which is higher than the temperature needed to excite the  CO(2-1) line (16.6K). Fortunately at high redshift, ALMA can only observe higher excitation lines, which are less affected. At $z>6$, only CO(6-5) ($T_e$=116.16K) and higher excitations can be observed with ALMA. The relation would be still valid at lower redshift ($z\lessapprox5$). However, the effect of the higher CMB temperature need to be examined carefully at higher redshifts. 

 Although we did not find any sign of redshift evolution, we also note that it is expected the gas fraction in a galaxy should increase as a function of redshift. Therefore, it is important to repeat this test when larger sample become available in the future.

\subsection{Estimating gravitational magnification using $L'_{CO}/\mu$-FWHM relation}

If we assume the $L'_{CO}/\mu$-FWHM relation, the equation \ref{bestfit} can be used to estimate $\mu$. We re-write the equation as follows.

\begin{equation}\label{eqmu}
\mu  = L'_{CO} \times(FWHM/1.3)^{-4.1} 
\end{equation}

The units for  $L'_{CO}$ and FWHM are  K Km s$^{-1}$ pc$^2$ and Km s$^{-1}$, respectively.
The equation can be useful when, for example, a submm galaxy is detected behind a galaxy cluster, and subject to gravitational magnification, but the detailed lens model is not available. Using the equation \ref{eqmu}, one can estimate $\mu$ and thereby intrinsic luminosities of the submm galaxy.

\section{Discussion} \label{sec:cosmology}

\subsection{A possible cosmological application}

If the correlation is independent of redshift, the luminosity distance to CO emitting galaxies can in principle be measured by estimating $L'_{CO}$ from FWHM. Such a distance measure would be useful because  already now, galaxies at z$\sim$6 have been detected in CO with ALMA \citep{2013Natur.496..329R,2013ApJ...773...44W}.  This method can be used up to such a high redshift, where it is difficult for other distance measures to reach. Here we assess whether such an approach could be feasible.

$L'_{CO}$ in Figs 1 and 2 assumes a cosmology;  $(\Omega_M,\Omega_\Lambda, h)$=(0.3,0.7,0.7). Therefore, we need to calibrate the correlation by using CO detected galaxies with a measured distance from another method. This, however, is difficult because CO detected galaxies are mostly at very high redshifts, where other distance measures cannot easily reach. 

There are a number of ways to calibrate the distances, but here we use a method often used to calibrate gamma ray burst.
We use an empirical formula for the luminosity distance as a function of redshift based on 192 type Ia supernovae at 0.168$<$z$<$1.755  ($D_L/10^{27}cm=6.96 z^{1.79}+14.79 z^{1.02}$; equation 1 of \citet{2009JCAP...08..015T}. This empirical formula does not assume any cosmology, but just that the type Ia supernovae are standard candles. We use the formula to assign luminosity distances to our submm galaxies at z$\leq$3, and thus to calibrate our $L'_{CO}$-FWHM correlation. Then, we apply the $L'_{CO}$ -FWHM correlation to measure luminosity distances to the z $>$ 3 submm galaxies. 

The results are shown in Fig.3. Along with the data points, we plot three different cosmological models with $(\Omega_M,\Omega_\Lambda)$=(0.3,0.7), (0.3,0.0), and (1.0,0.0) in the black solid, green dot
ted and blue-dashed lines, respectively. At this stage, it is clear that the scatter is 
more than an order of magnitude larger than needed to obtain meaningful constraints on the cosmological parameters.
%too large to obtain tight constraints on cosmological parameters. 
%However, Fig.3 shows a potential of the methodology to extend the Hubble diagram to much higher redshifts.

\subsection{Sources of scatter in the relation, and prospects for reducing it}

%In the last sub-section, we attempted to use the $L'_{CO}$-FWHM correlation to measure cosmological distance. However, the scatter was more than an order --- no where near to perform a precision cosmology. Below, we discuss possible sources of uncertainties in more detail.

The scatter around the $L'_{CO}$-FWHM correlation (43\%) is consistent with what is expected from randomly oriented discs, so if most submm galaxies are rotating spiral-like galaxies, then the largest uncertainty likely comes from the unknown inclinations. 
%In the future, the inclination effect can be handled in the following two ways; (i) Inclination can be measured in the near future. 
With the full operational array, ALMA’s spatial resolution will be as good as 0.037” at 110 GHz (for CO observation) in the extended configuration. At z$\sim$6, this corresponds to 0.2 kpc, which is sufficient to resolve submm galaxies, which typically extend over several kpc \citep{2013ApJ...773...44W,2014A&A...565A..59D}. 
With such observations it will also be possible to exclude 
mergers and pressure-supported galaxies which may be causing some of the extreme outliers from the relation, and 
to select 
samples of rotationally dominated galaxies.
%pure, rotational galaxies with ALMA.  (ii)
Furthermore, because the inclination is a random effect and its distribution is known (with $<$sin~i$>$$\sim$0.79 and 1$\sigma\sim$0.22), one can statistically correct for the effect. For the Tully-Fisher relation, for example, it has been shown that even when no information on the inclination is available, by using maximum likelihood estimation, the true relation can be recovered with only 1.5 times larger statistical error than when inclinations are known with zero uncertainty \citep{2013ApJ...777..140O}.
As a reference, using the Tully-Fisher relation, \citet{2000ApJ...529..698S} obtained $H_0$=71$\pm$4 (random) $\pm$7 (systematic), i.e., 6 and 10\% of statistical and systematic errors, respectively.

Another possible source of scatter is calibration errors, as even state of the art sub-mm telescopes can only be calibrated to 10\% at best for individual observations. Such errors are also expected to improve drastically as large, homogeneous statistical samples become available from ALMA, as the error on the median reduces with $\sqrt{N}$. In a sample of 10000 submm galaxies for example, the statistical error on the median will reduce to ~1\%.

%Another concern is that the state-of-the-art submm telescopes today can only be calibrated to $\sim$ 10\% at best. However, this is for individual objects. Once a statistical sample of submm galaxies is obtained, the statistical errors on the median reduces with $\sqrt{N}$. If $>$10,000 submm galaxies become available with ALMA, the statistical errors reduces to 1\%.
%In the near future, the superb sensitivity of ALMA enlarges the sample of CO galaxies. Therefore, a progress in statistics can be expected.
Note that this assumes the calibration error is random. If there is a systematic calibration error we are unaware of, this error will be added to the distance measurement.

%ALMA's high spatial-resolution also reduces uncertainties associated with $\mu$. A large number of sources will also statistically reduce the effect of  $\mu$. Because of flux conservation in gravitational lensing, the average magnification of a large number sources approaches unity \citep{2005ApJ...631..678H,2013PhRvD..88d3511W}. 

Another possible source of scatter in the relation which will be brought down in future ALMA surveys, is conversions between CO transitions. In such samples it will be possible to study galaxies in the same transition as a function of redshift, or subsets of galaxies with well-constrained transition conversions.
Once the observational and systematic errors have been brought down sufficiently to measure the intrinsic scatter of the relation it will to asses its possible cosmological applications. At the moment, it is impossible to tell.

%In addition to these, the uncertainty in conversions between different CO transitions need to be carefully treated. It is strait-forward to observe galaxies in the same CO transition across the redshift range. But this is difficult within the limited frequency range of a single telescope. Alternatively, one can limit the sample to galaxies with well-known transition conversion. In either case, we need to wait for higher quality data in the future.

%When such ALMA data become available, the method proposed in this work will be instrumental for measuring luminosity distances to extremely high-redshift. Already now, galaxies at z$\sim$6 have been detected in CO with ALMA \citep{2013Natur.496..329R,2013ApJ...773...44W}.  This method can be used at such a high redshift and beyond, where it is difficult for other distance measures to reach. Information on the expansion of the Universe during this early, previously unexplored era, will help us constrain the properties of dark energy, in particular, whether it varies over time or not.

%\section{Discussion}\label{sec:discussion}
%\label{discussion} 

%\section{Summary}

%---------------------------------------------------------- figure table ----------------------------------------------------------

\begin{acknowledgements}
We greatly appreciate the anonymous referee, whose insightful comments helped improving the paper significantly.
We acknowledge the support by the Ministry of Science and Technology of Taiwan through grant NSC 103-2112-M-007-002-MY3. The Dark Cosmology Centre is funded by the Danish National Research Foundation.  
\end{acknowledgements}

%-------------------------------------------------------------------

\bibliography{201404_goto} 
\bibliographystyle{aa}

\end{document}